\documentclass[10pt,letterpaper]{article}
\usepackage[T1]{fontenc}
\usepackage[utf8]{inputenc}
\usepackage{lmodern}

\usepackage[letterpaper,textwidth=6.5in,textheight=9in,top=1in,
  headheight=14pt,headsep=25pt,footskip=30pt]{geometry}
\usepackage{amsmath,amssymb,amsthm}
\usepackage{booktabs,tabularx,longtable,array}
\usepackage{graphicx,xcolor,tikz}
\usetikzlibrary{arrows.meta,positioning}
\usepackage{listings}
\usepackage{microtype}
\usepackage{enumitem}
\usepackage{fancyhdr}
\usepackage[hidelinks]{hyperref}
\hypersetup{pdftitle={Rózsa: Certified Equational Reasoning},pdfauthor={Georgios Kissas},pdfsubject={SAIR equational implication solver system description},pdfkeywords={magma identities, equational reasoning, Lean certificates, finite countermodels}}
\usepackage{caption}
\usepackage{needspace}
\setlist[itemize]{topsep=3pt,itemsep=2pt,leftmargin=*}
\setlist[enumerate]{topsep=3pt,itemsep=2pt,leftmargin=*}
\newcommand{\op}{\mathbin{\diamond}}
\newcommand{\sat}{\models}
\newcommand{\code}[1]{\texttt{\detokenize{#1}}}
\newcommand{\sys}{R\'{o}zsa}
\newcommand{\E}[1]{\textsf{E#1}}
\newtheorem{example}{Example}
\definecolor{ink}{RGB}{29,48,68}
\definecolor{wash}{RGB}{244,247,249}
\fancypagestyle{plain}{%
  \fancyhf{}\fancyfoot[C]{\thepage}}
\makeatletter
\AtBeginDocument{\let\rozsaRunningTitle\@title}
\renewcommand{\section}{\@startsection{section}{1}{\z@}%
  {-2ex plus -0.5ex minus -0.2ex}{1.2ex plus 0.2ex}%
  {\large\bfseries\raggedright}}
\renewcommand{\subsection}{\@startsection{subsection}{2}{\z@}%
  {-1.8ex plus -0.5ex minus -0.2ex}{0.8ex plus 0.2ex}%
  {\normalsize\bfseries\raggedright}}
\renewcommand{\subsubsection}{\@startsection{subsubsection}{3}{\z@}%
  {-1.5ex plus -0.5ex minus -0.2ex}{0.5ex plus 0.2ex}%
  {\normalsize\bfseries\raggedright}}
\renewcommand{\@maketitle}{%
  \vspace*{-1.5em}%
  \begin{center}
    \rule{\linewidth}{1.5pt}\par
    \vspace{0.6em}
    {\LARGE\bfseries\@title\par}
    \vspace{0.4em}
    \rule{\linewidth}{0.6pt}\par
    \vspace{0.8em}
    {\normalsize\@author\par}
    \vspace{0.5em}
    {\small\@date\par}
  \end{center}
  \vspace{0.5em}}
\makeatother
\renewenvironment{abstract}{%
  \begin{center}\large\bfseries\scshape\abstractname\end{center}%
  \begin{quote}\small}{\end{quote}\vspace{0.5em}}

\title{\sys: Certified Equational Reasoning}
\author{Georgios Kissas\\Swiss Data Science Center, ETH Zurich}

\begin{document}
\maketitle

\begin{abstract}
When does one equation imply another? Stage 2 of the SAIR Mathematics Distillation Challenge required teams to submit solvers to answer this question for sets equipped with a binary operation. For these sets, associativity, commutativity, or any other algebraic property was not assumed. Each answer required a proof or counterexample that could be verified by Lean. Programs also had to respect fixed resource limits. We present \sys{}, a Python solver that searches for both kinds of evidence. It begins with simple tests and considers more extensive searches when those fail. To find a proof, it applies a given equation and derives any further equations as needed, keeping a record of each step. To find a counterexample, it searches for operations on finite sets. It also tests one fixed infinite model. The given equation must hold for every assignment, while the target equation must fail for at least one. In either case, \sys{} turns the evidence into a Lean certificate. The submitted solver solved 199/200 problems (99.5\%) in the official main evaluation getting the 6th overall position. Additional tests show that the solver can prove many implications linked by long chains of known results, but has limited success with infinite counterexamples. We explain the search methods, how they fit together, and what the unsolved cases reveal about their limits.
\end{abstract}

\section{Introduction}
\label{sec:intro}
An equation such as $x\op x=x$ is easy to write, but its properties can be hard to discover. A \emph{magma} is a set with a binary operation $\op$: a rule that takes two elements and returns an element of the same set. We assume neither associativity nor commutativity. Given a starting law $H$, called the \emph{hypothesis}, and a law $T$, called the \emph{target}, we ask whether $T$ holds in every magma that satisfies $H$. A solver must justify either answer: derive $T$ from $H$, or give a set and operation for which $H$ holds and $T$ fails.
The Equational Theories Project organized formally checked implications between 4,694 small magma laws \cite{etp}. The SAIR Mathematics Distillation Challenge, Stage 2, asked submitted programs to solve problems of this kind. Each answer needed a \emph{certificate}, a proof written in Lean \cite{lean4}. The challenge uses a \emph{judge} that creates a proposition for the chosen answer and checks that the submitted code proves it.

\sys{} separates the search for an answer from the correctness checks. For each  proof, it records the equations and steps used in the derivation. For each counterexample, it records the set, its operation, and an assignment where the target fails. Then it expresses this evidence in Lean code. Simpler strategies are run first, and the solver considers more extensive searches for harder cases.
The methods complement each other. Proving an intermediate law can make a complicated target easy to derive. Searching formulas with a few coefficients can find a counterexample without examining every operation table. The same derivation machinery can prove the target equation or establish that an infinite model satisfies the hypothesis. The paper explains these connections and the reasons for the search order.

\paragraph{Relation to existing systems.}
The proving methods are based on \emph{rewriting}, which applies equations to parts of expressions, and \emph{completion}, which derives new equations to use as rules \cite{knuthbendix,huet}. Recording completion also keeps the origin of each rule so that a proof can be reconstructed \cite{recording}. Counterexample search draws on finite model finding \cite{mace4} and methods that repair violated constraints \cite{minconflicts}. Experiments with Vampire provide a broader reference for proving implications in the Equational Theories Project \cite{janota}.
Two recent systems offer useful comparisons. Janota, Rawson, and Schulz extract models, possibly infinite, from fully processed equation systems. They certify termination and confluence for 261 resulting rewrite systems: rewriting stops, and competing reductions can be joined \cite{saturationmodels}. Those certificates establish properties of the rewrite systems. SAIR acceptance requires a Lean certificate for the requested implication. Another SAIR solver uses ordered unit superposition, a method for deriving consequences of equations, together with systematic finite-model search and several infinite constructions \cite{ma}. \sys{} uses several bounded proof searches. Its counterexample methods include repairing finite tables, filling their entries subject to constraints, and testing one infinite construction and its reverse. 

The presentation follows a problem through the system. Section \ref{sec:representation} explains its mathematical meaning and tree representation. Section \ref{sec:architecture} explains the schedule. Sections \ref{sec:true} and \ref{sec:false} describe proof and countermodel discovery, Section \ref{sec:certificates} shows how both become checked certificates. Execution limits follow in Section \ref{sec:resources}. Results, reproducibility, and discussion complete the main text. Exact parameters, additional implementation details, and post-evaluation failure analysis are in the appendices.

\paragraph{The implementation being described.}
The paper describes the mathematical search methods in the unchanged 175,375-byte submission. Appendix \ref{app:implementation} gives the hash that identifies this file. Appendix \ref{app:optionalproposal} describes the optional language-model fallback, and Section \ref{sec:evaluation} states which features were enabled in the measured runs.

\section{What the problem means and how expressions are stored}
\label{sec:representation}
\subsection{What a true or false answer means}
A magma is a set $G$, called its \emph{carrier}, together with a function $\op:G\times G\to G$. A \emph{term} is an expression built from variables and this operation. An \emph{identity} $s=t$ is an equation that holds for every assignment of its variables to elements of the set. Write $G\sat H$ when the identity $H$ holds in $G$. Saying that $H$ implies $T$ means
\[
H\sat T\quad\Longleftrightarrow\quad
\text{for every magma }G,\; G\sat H\ \Longrightarrow\ G\sat T.
\]
A \emph{model} of an identity is a set and operation in which it holds. To disprove an implication, a certificate supplies a \emph{countermodel}: a magma $G$ satisfying $H$, together with one assignment $\rho$ for which $T$ fails. The two checks are different. The hypothesis must hold for every assignment; one target violation is enough.

Birkhoff's completeness theorem connects proofs and models: $H\sat T$ holds exactly when a finite sequence of equational reasoning steps proves $T$ from $H$ \cite{birkhoff,recording}. These steps substitute expressions for variables, use $s=s$, reverse an equality, chain equalities, or replace a subexpression by an equal one. Their technical names are instantiation, reflexivity, symmetry, transitivity, and congruence. The proving methods record these steps.
If there is no derivation, there is a countermodel. One construction groups expressions whose equality follows from $H$ and treats each group as an element. The operation combines representatives of the groups; congruence ensures that the resulting group does not depend on the representatives chosen. This \emph{quotient term magma} satisfies $H$ but keeps the target expressions distinct. It can be infinite. Birkhoff's theorem explains why a proof or a countermodel exists; it does not guarantee that \sys{} will find one within its search limits. The challenge permits both finite and infinite countermodels. The proposition for a false answer allows any carrier type and magma, with no \code{Finite} or \code{Fintype} requirement. A certificate for either kind of model therefore proves the same kind of claim. The \emph{order} of an equation is the number of operation symbols on its two sides \cite{etp}. It does not count elements in a model. An order-five equation does not need to have a five-element countermodel. We use \emph{carrier size} to mean the number of elements in a finite model.

\begin{example}[Two distinct kinds of evidence]
Idempotence $H:x\op x=x$ does not imply commutativity $T:x\op y=y\op x$. On $\{0,1\}$, define $a\op b=a$. The hypothesis holds for both elements, while $0\op1=0$ and $1\op0=1$. For an example of a true implication, the projection hypothesis $a\op b=a$ implies $((x\op y)\op z)\op w=x$, by three applications of the hypothesis. The first problem needs a model; the second needs a derivation.
\end{example}

\subsection{Storing expression trees and keeping variable order}
The parser turns each expression into a binary tree. A variable is a leaf; an operation joins two subtrees. Thus $(x\op y)\op z$ and $x\op(y\op z)$ remain distinct. Internally the nodes are tuples: \code{("v", name)} and \code{("o", left, right)}.
The ASCII marker \code{*} and the diamond notation denote the same operation. Unparenthesized sequences are parsed left-associatively, while parentheses preserve their explicit grouping. Malformed input raises an error. Preserving the exact tree is essential because the solver cannot assume associativity. When an expression occurs repeatedly, the solver stores it once and reuses it, a technique called \emph{interning}. It also saves the expression's printed form, size, depth, and variables. Size counts all tree nodes, including variable leaves, rather than just operation symbols. Search uses these properties repeatedly, so saving them avoids repeated work.
Variable names are collected in the order of their first occurrence across the equation. The solver uses that same order when passing arguments to the hypothesis in Lean. The tree preserves the equation's meaning; variable order makes the generated proof apply the hypothesis correctly.

\subsection{Shared operations and their roles}
The searches use three basic operations. \emph{Matching} asks whether a term fits a pattern whose variables are placeholders. \emph{Substitution} fills those placeholders with terms. \emph{Unification} allows variables on both sides to change and seeks a substitution making the terms identical. An occurs check excludes cyclic assignments such as $x\mapsto x\op y$.
For example, matching $x\op y$ against $(a\op b)\op c$ gives $x\mapsto a\op b$ and $y\mapsto c$. Unifying $x\op a$ with $b\op y$ can instead give $x\mapsto b$, $y\mapsto a$. The distinction explains why ordinary rewriting uses matching, whereas overlapping two rules uses unification. Two more operations support the rule overlaps used later. \emph{Context replacement} changes one subtree and leaves its surroundings intact. \emph{Variable standardization} gives two copies of a rule separate variable names, so that variables are not accidentally treated as the same before unification. Simple features of the expression trees select promising early proof methods, described in Section \ref{sec:true}. These features choose where to search; only a derivation can establish the answer. Some model searches also test an operation with its arguments reversed. This is the \emph{opposite operation}, $a\op^{\mathrm{op}}b=b\op a$. To evaluate a term this way, the solver swaps the two children at every node of its tree, a transformation called \emph{mirroring}. It can therefore test both an operation and its reverse using the same evaluation code.

\section{How the search methods are combined}
\label{sec:architecture}
\subsection{Several searches, one checking procedure}
The solver tries several methods in a fixed sequence. We call this collection the mathematical \emph{portfolio}. Each method receives the same hypothesis and target. When a method finds a possible answer, the solver builds a Lean certificate. It checks the certificate's format and size before sending it to the judge. If the judge accepts it, the run ends. Otherwise, the solver moves to the next method. Finishing a search without an answer leaves the problem undecided. All methods can use the parsed expressions, cached computations, remaining time, and summaries of earlier searches. Most nevertheless maintain their own equations and queues. A method may reuse a lemma during its own search, but the next method usually starts afresh. The stages are connected by the evidence they produce, not by a shared queue of unfinished work. This keeps their interaction simple, but later stages do not continue an earlier stage's unfinished search. The schedule alternates proof and countermodel searches, although the next two sections explain them separately. Table \ref{tab:cascade} shows each group's role before those sections describe the individual methods. Appendix \ref{app:parameters} gives the exact order and limits. The early projection search also has an entry condition based on expression structure, specified in Appendix \ref{app:conversionbounds}.

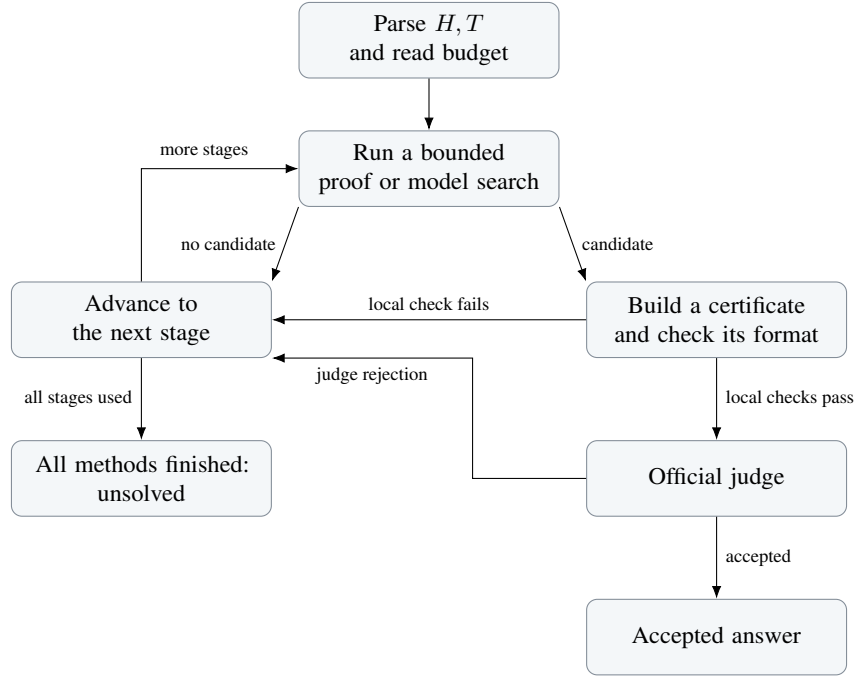
\begin{figure}[t]
\centering
\begin{tikzpicture}[>=Latex,
 box/.style={draw=ink!55,rounded corners,fill=wash,align=center,font=\small,minimum height=10mm,text width=32mm}]
\node[box] (input) at (0,0) {Parse $H,T$\\and read budget};
\node[box] (search) at (0,-1.7) {Run a bounded\\proof or model search};
\node[box] (next) at (-3.8,-3.7) {Advance to\\the next stage};
\node[box] (emit) at (3.8,-3.7) {Build a certificate\\and check its format};
\node[box] (judge) at (3.8,-5.8) {Official judge};
\node[box] (unsolved) at (-3.8,-5.8) {All methods finished:\\unsolved};
\node[box] (done) at (3.8,-7.9) {Accepted answer};
\draw[->] (input)--(search);
\draw[->] (search.south east)--node[right,font=\scriptsize]{candidate}(emit.north west);
\draw[->] (search.south west)--node[left,font=\scriptsize]{no candidate}(next.north east);
\draw[->] (emit.west)--node[above,font=\scriptsize]{local check fails}(next.east);
\draw[->] (emit)--node[right,font=\scriptsize]{local checks pass}(judge);
\draw[->] (judge)--node[right,font=\scriptsize]{accepted}(done);
\draw[->] (next.north)|-node[pos=.7,above,font=\scriptsize]{more stages}(search.west);
\draw[->] (next)--node[left,font=\scriptsize]{all stages used}(unsolved);
\draw[->] (judge.west)--++(-1.5,0)|-node[pos=.75,below,font=\scriptsize]{judge rejection}(next.south east);
\end{tikzpicture}
\caption{Every search method follows the same checking procedure. The solver advances when search finds nothing or a candidate fails a check. Entry conditions can skip a method. Methods run one after another, not concurrently, and a run can finish unsolved.}
\label{fig:architecture}
\end{figure}

\begin{table}[t]
\centering\small
\begin{tabularx}{\linewidth}{@{}r p{.32\linewidth} X@{}}
\toprule
Step & Search group & Why it occurs here \\
\midrule
1 & Small operation tables & Find simple counterexamples before building large proof searches. \\
2 & Direct applications and short proofs & Settle easy true cases by applying the hypothesis or following a short chain. \\
3 & Random and larger structured tables & Try more finite operations after the cheapest searches fail. \\
4 & Formula-based models and derived proofs & Try bit-vector formulas and simple intermediate laws before larger searches. \\
5 & New-rule search and matching target parts & Derive useful equations and use the target's shared expression structure. \\
6 & Broader rule and rewrite searches & Use equations that are poor simplifiers, then search beyond short rewrite paths. \\
7 & Richer formulas and table repair & Try operations outside the earlier formula families and repair less regular tables. \\
8 & Specialized final searches & Try an infinite counterexample, fill a six-element table by constraints, or prove a law that selects one operand. \\
\bottomrule
\end{tabularx}
\caption{Overview of the search order. Some rows group several consecutive methods. Modular affine search runs within step~6, before the deeper rewrite searches. Appendix \ref{app:order} gives the exact order.}
\label{tab:cascade}
\end{table}

\subsection{Why this order?}
Quick methods run first because a direct application of the hypothesis or a small counterexample can settle a case before a large search begins. The schedule then alternates between looking for a proof and looking for a counterexample. Either search can be unproductive when the opposite answer is true, so neither receives the whole budget at the outset.
During development, new methods were placed after an early sequence that had already passed the existing tests. This kept the quick methods that were already successful in place. Some later searches first try to prove a stronger law, such as a projection law, from which the target follows directly. The design also accounts for \emph{checking} cost. Finding a mathematical answer is not enough: its certificate must fit the size limit and finish Lean elaboration, the process that turns proof code into checked proof terms. Model searches therefore favor manageable finite sets, formulas can replace large tables, and generated proofs share intermediate lemmas. Time to find an answer and time to obtain an accepted certificate are different measurements.
These considerations explain the deployed schedule. Section \ref{sec:discussion} assesses its behavior, while Appendix \ref{app:failures} analyzes the searches left unfinished after evaluation.

\section{Producing proofs of true implications}
\label{sec:true}
The proof branch tries three progressively broader ideas: use the hypothesis directly, search through its instances, then derive new identities when direct rewriting stalls. Every successful route records the equalities needed for its certificate.

\subsection{Direct substitution and total collapse}
The simplest proof is an instance of the hypothesis. The solver looks for one substitution that turns both sides of the hypothesis into the corresponding sides of the target. If it finds one, the Lean proof applies \code{hyp}, the name for the input hypothesis, to those terms. If the sides match in reverse order, the proof also reverses the equality. A target whose sides are already identical needs only reflexivity. The \emph{collapse} method recognizes $x=u$, or $u=x$, where $x$ does not occur in $u$. Hold all other variables fixed and assign $x$ first to $a$ and then to $b$. The hypothesis gives $a=u=b$, so any two elements are equal. Every target identity then follows. The solver expresses this argument as a Lean lemma and applies it to the two sides of the target; it does not assume the collapse without proving it.

\subsection{Searching for a sequence of rewrites}
The solver first looks for a sequence of equation applications that connects the target's two sides. It treats expressions as vertices in a graph. Each edge applies the hypothesis at one position in an expression, in either direction. This is \emph{recorded conversion}: searching for a rewrite path while keeping its justification. Matching one side of the hypothesis can leave some variables unspecified. A finite pool of variables and subexpressions supplies their possible values. The search limits this pool, the number of expressions generated at each step, the size and depth of those expressions, and the total number of visited states.
The pool contains target variables and subexpressions of the target and hypothesis. It keeps only expressions using target variables, then sorts them by size and stored representation. This gives concrete choices for missing substitutions without allowing unlimited branching. Path depth counts successive rewrites; tree depth counts nested operation nodes. They are separate limits.
These restrictions can exclude a necessary intermediate step. The method is therefore incomplete even if its time allowance is increased. Birkhoff's theorem guarantees a finite proof for a true implication, but this bounded search need not visit that proof.

Each edge stores a Lean proof of the equality it uses. A rewrite of the whole expression applies the hypothesis directly. A rewrite inside a larger expression uses \code{congrArg}, Lean's rule that equal arguments have equal function values. Reversing an equality uses \code{Eq.symm}. Each visited expression records its predecessor, so a successful path supplies the complete equality chain.
The first breadth-first search explores paths in increasing length, starting from the target's left side. A later search allows longer paths. A reverse search starts from the right side and favors rewrites that shrink expressions. A guided search favors expressions resembling the destination. These variants explore different parts of the same graph. Appendix \ref{app:conversionbounds} gives their path and state limits.

\Needspace{10\baselineskip}
\begin{example}[A certificate is a recorded calculation]
With associativity as the hypothesis, the path
\[
((a\op b)\op c)\op d
= (a\op b)\op(c\op d)
= a\op(b\op(c\op d))
\]
uses two hypothesis instances. The solver stores the substitutions and each edge equality. It then emits the same chain as a Lean \code{calc} proof. This illustrative example explains the mechanism; it is separate from the benchmark measurements.
\end{example}

\subsection{Proving equality by matching corresponding parts}
Sometimes the two target expressions already have the same outer shape. The solver can then prove equality of their corresponding parts instead of searching through whole expressions. If $s_1=t_1$ and $s_2=t_2$ have proofs, congruence gives $s_1\op s_2=t_1\op t_2$. This recursive method is called \emph{structural conversion}.
One variant matches one side of the hypothesis and then checks what the same substitution gives on the other side. A second variant, \emph{joint structural conversion}, matches both sides together. For $H:x\op y=y\op z$, matching $x\op y$ against $a\op b$ determines $x=a,y=b$ but leaves $z$ unspecified. Matching the other side against $b\op(a\op a)$ determines $z=a\op a$. Joint matching can also make this application inside a shared surrounding expression, using congruence to justify the surrounding equality. Both variants build explicit proofs from congruence and transitivity, with limits on recursion and output size.

\subsection{Projection lemmas as reusable explanations}
A \emph{projection operation} always returns one operand: left projection is $a\op b=a$, and right projection is $a\op b=b$. Under left projection, every term evaluates to its leftmost variable; under right projection, to its rightmost variable. A complex target therefore becomes immediate if both sides have the same relevant endpoint variable.
The solver checks the target's leftmost or rightmost variables before trying to prove a projection law from $H$. If the relevant variables agree, that law would be enough to prove the target. The variable check only selects the intermediate goal; the solver still has to derive the projection law from the hypothesis.
It searches for that derivation using rewriting, overlaps, and the completion methods described next. Once the law is proved, both target expressions reduce to the same variable. One intermediate result can thus replace a much longer sequence of rewrites.

\subsection{Deriving a law from overlapping copies of the hypothesis}
The same equation can apply to both a whole expression and one of its parts. If the two applications start from the same expression, their results must be equal. To find such a \emph{self-overlap}, the solver makes a renamed copy of the hypothesis and unifies one rule occurrence with a nonvariable subexpression of the other. This produces a common starting expression and two valid reductions.
The early methods keep up to sixteen distinct equations obtained this way, then search using the hypothesis and one new equation. Another method checks whether simplifying these equations reveals a projection law. This can find a useful first lemma before attempting a much larger search for consequences. Giving the copies separate variable names avoids accidental identifications. The stored proof records both applications and the expressions around them.

\Needspace{18\baselineskip}
\begin{example}[A running overlap-to-projection explanation]
\label{ex:running}
Consider the illustrative input
\[
H:\ x\op((y\op z)\op w)=x,
\qquad T:\ ((a\op b)\op c)\op d=a.
\]
The shared expression $A=x\op((y\op z)\op((u\op v)\op w))$ has two reductions. At its root, $H$ gives $A=x$. Inside its right operand, $H$ with first argument $y\op z$ gives $A=x\op(y\op z)$. Combining those equalities derives $K:x\op(y\op z)=x$. Now repeat the two-path argument using $K$. The expression $B=x\op(y\op(z\op w))$ reduces to $x$ by $K$ at the root and to $x\op y$ by $K$ inside the right operand. Thus $P:x\op y=x$ follows. Applying $P$ three times proves $T$. The example shows how the methods connect. The expression tree identifies where a rule can apply. Two applications give the intermediate equation $K$, and another pair gives the projection law $P$. That law settles the target. Section \ref{sec:certificates} shows how the recorded steps become a Lean certificate.
\end{example}

\subsection{Deriving new rules and recording their proofs}
Rewriting uses equations already known. \emph{Completion} derives further equations and then uses them as rules too. A consequence found in one round can therefore help produce another in a later round. Classical completion aims to build a terminating, confluent system: rewrites stop, and competing paths can be joined \cite{knuthbendix,huet}. \emph{Recording completion} also preserves how each rule follows from the original equations \cite{recording}. \sys{} uses a bounded version to search for a proof, without requiring completion of the whole system. The key step is the overlap just illustrated, now applied to any two available rules. Let $\ell_1=r_1$ be the outer rule and $\ell_2=r_2$ the inner rule. Give their variables separate names, then unify the subexpression at a nonvariable tree position $p$ in $\ell_1$ with $\ell_2$. The resulting substitution $\sigma$ gives a common starting expression $c=\ell_1\sigma$. Applying the outer rule gives $r_1\sigma$. Applying the inner rule at position $p$ gives $c[p\leftarrow r_2\sigma]$, where the brackets mean replacement of that subtree. The equation between the two results is a \emph{critical pair}. Its proof reverses one reduction and follows the other. All these steps are recorded in the proof graph described in Section \ref{sec:certificates}. The algorithm simplifies each resulting equation, discards equations with identical sides and duplicates, and adds useful new rules. It checks whether a new equation directly matches the target. It also tries to \emph{normalize} both target sides: apply selected rewrite rules repeatedly until no rule applies or a bound is reached. If both sides reach the same expression, the recorded steps prove the target. Only these two paths need to meet. There is no need to show that rewriting gives every expression a unique final form.

A consequence's \emph{generation} counts its stage of derivation. The original hypothesis has generation zero; a new consequence has one plus the maximum generation of its parent rules. The core loop is:
\begin{enumerate}
\item Record the hypothesis with the substitution that leaves its variables unchanged, and choose a direction for rewriting.
\item Pair each new rule with each existing rule. Give their variables separate names and find nonvariable overlaps, trying each rule as the outer rule.
\item Record both reductions and queue the resulting equations that fit the term and generation limits.
\item Test whether the current rules can rewrite both target sides to the same expression.
\item Select a queued equation, simplify it while recording the rewrites, test whether it matches the target, and add it if it is a new consequence.
\item Repeat until a proof is found, a bound is reached, or the queue is empty.
\end{enumerate}
In Example \ref{ex:running}, $K$ is a first-generation consequence and $P$ is a later consequence using $K$. The proof graph retains both overlap explanations even after their equations become rewriting tools. Seven completion configurations vary the search limits and the preference for small or shallow expressions. The ranking selects directions for rewriting; it is a search heuristic, not part of the proof. The stored equality can justify a rewrite in either direction. Appendices \ref{app:parameters} and \ref{app:implementation} give the exact configurations and explain their source-code labels.
The \emph{target-interleaved} configuration alternates between generating overlaps and examining equations already in the queue. It gives priority to small equations that resemble the target, so potentially useful equations are considered before more work accumulates. Resemblance is measured by comparing printed symbols; the score does not establish mathematical usefulness. Appendix \ref{app:conversionbounds} gives the formula.

\subsection{Two ways to control the cost of completion}
An equation may help discover an overlap even if it is a poor rule for simplifying expressions. \emph{Active completion} therefore keeps separate sets of \emph{reducers}, used for simplification, and \emph{generators}, used for overlaps. A reducer must pass tests on variable occurrences, size, and depth. If neither direction of an equation passes these tests, the equation can still generate overlaps in both directions.
If a new reducer simplifies an existing equation, the solver retires the old entry and inserts its simpler replacement. This \emph{backward simplification} avoids redundant work without losing the recorded proof. Appendix \ref{app:conversionbounds} states the tests precisely. Keeping reducers separate also avoids trying every equation in both directions during normalization. The ranking guides search, while explicit bounds limit its cost; it does not guarantee that rewriting always terminates. Appendix \ref{app:implementation} gives a concrete reason.

\emph{Lazy projection completion} avoids building full proofs for equations that may never be used. For each overlap, it initially stores only the two rules, the position, and the substitution. It builds the derivation when the resulting equation is selected from the queue. Most selections favor equations resembling the target; every fifth instead takes one of the oldest pending overlaps. The aim is to prove a projection law and then apply it to the target. Active completion decides which equations to use for simplification; lazy completion delays the work of building their proofs.

\section{Discovering countermodels for false implications}
\label{sec:false}
To disprove an implication, the solver needs a set and an operation for which $H$ holds but $T$ fails. Some methods search families of operations defined by formulas with a few parameters. Others work directly with finite operation tables, either changing complete tables or filling unknown entries subject to constraints. The solver also tests one fixed infinite model. The schedule in Section \ref{sec:architecture} interleaves these methods with proof search. Each explores a different class of possible counterexamples; searching larger finite tables cannot find one that requires an infinite set.

\subsection{Searching small operation tables}
A finite magma on $n$ elements is an $n\times n$ table with entries in $\{0,\ldots,n-1\}$. Terms are evaluated recursively using table lookups. For every candidate, the local validator checks all $n^{v_H}$ assignments of the hypothesis's $v_H$ variables, and seeks a failing assignment among the target's $n^{v_T}$ possibilities.
There are $n^{n^2}$ possible operation tables. The initial exhaustive search enumerates the sixteen two-element tables and the 19,683 three-element tables, subject to its deadline. Exhausting these carriers excludes countermodels only at those sizes. Finite model finding can complement proof search, but a bounded absence of witnesses does not establish an implication \cite{mace4}. Before exhaustive enumeration, the solver tries tables built from simple formulas. These include constant operations, left and right projections, modular addition and subtraction, multiplication, minimum and maximum, and small affine formulas. One piecewise example is $x\op y=x$ when $x\ne0$ and $0\op y=y$. On a set of pairs, another construction is $(a,b)\op(c,d)=(a,d)$. Duplicate tables are removed within each carrier size. The first pass uses sizes two through five; a later pass tries these structured families at sizes six through nine.

\subsection{Affine formulas: searching coefficients instead of tables}
On the integers $0,\ldots,n-1$, with arithmetic modulo $n$, consider
\[
x\op y=ax+by+c\pmod n.
\]
An \emph{affine} operation is a weighted sum of its inputs plus a constant, here computed modulo $n$. Every term built from it is also affine. The solver represents that term by a \emph{signature}: one coefficient per variable and a constant. At an operation node, it multiplies the left coefficients by $a$, the right coefficients by $b$, and adds them. If the left and right subterms have constants $c_s$ and $c_t$, the new constant is $a c_s+b c_t+c$ modulo $n$.
For $n\geq2$, two terms define the same function exactly when their signatures agree. To see why, set all variables to zero to compare constants, then set one variable to one to compare its coefficient. A candidate is a countermodel if the two sides of $H$ have the same signature but the two sides of $T$ do not. The solver also checks the complete operation table before generating the certificate.
The modular search tries $10\leq n\leq17$, enumerating all coefficients and constants unless its short deadline is reached. Its certificate defines the arithmetic operation directly instead of including a large table. The formula therefore helps twice: it reduces the number of search parameters and gives a shorter description for Lean to check. The binary affine route applies the same idea to bit vectors: $\mathbb F_2^2$ and $\mathbb F_2^3$ have four and eight elements. Their addition is bitwise XOR. Operations have the form
\[
x\op y=\phi(x)\mathbin{\oplus}\psi(y)\mathbin{\oplus}c,
\]
where $\phi,\psi$ are linear maps, which preserve XOR, and $c$ is a fixed bit vector. Integers 0--3 or 0--7 encode the vectors. A linear map is determined by its values on the coordinate vectors, so the solver enumerates maps by choosing those values.
On four elements the search generates all such maps. On eight elements it uses only invertible maps, making the search smaller but excluding some operations. Each term's signature now contains a linear map for each variable and a constant. As in the modular-arithmetic search, the solver first checks whether the signatures satisfy the hypothesis, then validates the full table.

\subsection{Adding interactions to the formula families}
The next families introduce interactions that an affine operation cannot express. A \emph{Boolean linear form} is the parity of selected input bits. On four elements the quadratic route adds an output pattern when two such parity tests are both one:
\[
x\op y=L(x)\oplus R(y)\oplus c\oplus
v\,[\alpha(x)\land\beta(y)],
\]
Here $L,R$ are linear maps, $\alpha,\beta$ are nonzero Boolean linear forms, and $v$ is a nonzero output vector. The bracket is one when both tests pass, and zero otherwise. The added term is called rank-one bilinear: the two scalar tests control a single output pattern $v$. The gated route begins instead with an arbitrary unary map $u$ and changes its output only when the left input belongs to a selected set $S$ and the right input passes a parity test:
\[
x\op y=u(x)\oplus v\,[x\in S\ \land\ \beta(y)=1],
\]
The set $S$ is nonempty and proper, $\beta$ is a nonzero Boolean linear form, and $v$ is a nonzero output vector. The search also tries the opposite operation. Both families add cross-input dependence through a small number of parameters, rather than enumerating every four-element table. The switched construction uses a graph to generate tables. Choose a finite coordinate set and use its ordered pairs as vertices, with edges $(a,b)\to(b,c)$. Every pair of vertices $(a,b)$ and $(c,d)$ has a unique two-step path through $(b,c)$. Choosing that intermediate vertex as the operation's output gives $(a,b)\op(c,d)=(b,c)$.
To obtain less regular tables, the search replaces edges $u\to v$ and $s\to t$ with $u\to t$ and $s\to v$, excluding degenerate or already present replacements. It then checks whether the graph still gives exactly one intermediate vertex for every source--target pair. If so, those vertices define a new operation.
The submitted solver uses three choices per coordinate, giving nine elements. This graph search follows the ordinary nine-element structured tables in the extended pass. The graph property supplies a table, but does not by itself make it a counterexample: the solver must still check $H$ and $T$. These methods follow the broader use of linear and modified models in the Equational Theories Project \cite{etp}. Taken together, the formula and graph families trade freedom for a smaller search: they explore selected kinds of operation rather than every possible table.

\subsection{Building and repairing candidate tables}
The solver also generates random four- and five-element tables, using a seed derived from the problem identifier. Later searches try to repair tables on four, five, or six elements. They start with complete tables and record which entries were used whenever an assignment violates the hypothesis. The search changes one of those entries, tries its possible values, and favors changes that reduce the number of violations. It sometimes accepts a change that leaves the count unchanged, with a small fixed probability, to avoid getting stuck. Repeated starts and different seed streams explore different tables.
This follows the min-conflicts approach to constraint repair \cite{minconflicts}. The reason for choosing cells this way is simple: changing a cell used in a failing evaluation can repair that failure; changing an unrelated cell cannot. Once every hypothesis instance holds, the solver checks whether the target fails. If the target also holds, it uses the cells involved in a target evaluation to guide further changes. The basic repair method ranks changes by the number of assignments that violate the hypothesis. The \emph{joint local search} variant also uses target failure to break ties once the hypothesis holds. Both methods avoid evaluating the potentially larger set of target assignments while the hypothesis still has violations. They work first on the condition that must hold everywhere, before looking for a single assignment where the target fails.
The searches use fixed seeds and explicit limits. Appendix \ref{app:implementation} explains how seeds are chosen and which changes to a problem leave them unchanged. Fixed seeds determine the random-number sequences, but time cutoffs and set iteration can still change how far a run gets. The certificates can be checked independently regardless of how they were found.

\subsection{Filling a six-element table by constraints}
Local repair changes a complete table. A second method starts with unknown entries and rules out values that cannot satisfy the equations. This is a \emph{constraint satisfaction problem} (CSP). The submitted implementation uses six elements, so it searches for 36 table entries and the values of the subexpressions needed to check the equations. It looks for models with more elements than the exhaustive two- and three-element searches, while fixing the table size in advance.
A \emph{domain} records the possible values of an unknown, stored as a bit mask. \emph{Grounding} substitutes concrete elements for term variables. For example, grounding $x\op(x\op y)$ at $x=0,y=1$ introduces unknown results for $0\op1$ and $0\op(0\op1)$. For every hypothesis assignment, the values of the identity's two sides must agree; for one selected target assignment, the values of the target's two sides must differ. Function constraints connect these results to the operation table. The solver repeatedly removes values that conflict with the constraints. Unknowns required to be equal keep only the values they have in common. If two unknowns must differ and one has a fixed value, that value is removed from the other's domain. Each remaining value must have \emph{support}: it must fit at least one allowed combination of the other values and table entries in the constraint. For example, if an operation must return 2, a candidate left input is ruled out when no allowed right input can produce 2 with it. Once both inputs are fixed, the constraint also restricts the corresponding table entry. When no more values can be removed, the search chooses an unknown with the fewest possibilities and tries them in turn. Ties favor unknowns used by more constraints. A record of changes allows the search to undo a choice when it backtracks. Target witnesses are enumerated by their equality patterns. Assignments such as $(4,4,2,5)$ and $(0,0,1,2)$ share a pattern: the first two positions agree and the others are distinct. Relabelling the carrier makes one representative sufficient for each pattern. A restricted first pass requires the six diagonal outputs $a\op a$ to be distinct and tries their permutations in a fixed order. A second pass removes that requirement. This tries a smaller family before widening the search, with both passes sharing the same bounded allowance. Appendix \ref{app:parameters} gives the time and node limits.

\subsection{A fixed infinite counterexample construction}
Some implications hold for every finite magma yet fail on an infinite one. Finite table search cannot refute them. The solver therefore includes a fixed infinite carrier with a special element and two indexed chains:
\[
\{0\}\ \cup\ \{o_n:n\in\mathbb N\}\ \cup\ \{e_n:n\in\mathbb N\},
\]
The symbols $o_n,e_n$ name elements of the chains; they are not parity calculations on integers. A fixed piecewise operation moves along their indices. For example, $e_n\op0=o_{n+1}$ and $o_{n+1}\op0=e_n$. This is the \emph{ray} construction used throughout the paper. It satisfies the base law
\begin{equation}
x=(x\op y)\op((y\op y)\op z).
\label{eq:raylaw}
\end{equation}
Appendix \ref{app:ray} specifies the complete operation. To use this model, the solver must prove that the input hypothesis holds throughout the infinite set. It runs a short completion search to derive the hypothesis from~\eqref{eq:raylaw}. It also tests the opposite operation. In that case, it derives the mirrored hypothesis from the same base law, using the correspondence explained in Section \ref{sec:representation}. Finding an assignment where the target fails is not enough on its own: the hypothesis must also be proved.
The search for a target violation uses assignments from $\{0,o_0,e_0,o_1,e_1\}$. It recognizes different results only when they have different forms: $0$, $o_n$, or $e_n$. It misses unequal results of the same form, such as two $o$ elements with different indices. The Lean certificate defines the set and operation, proves the base law and the input hypothesis, and verifies the selected target violation. Thus the search tests only a few assignments, but the proof of the hypothesis covers the entire infinite set.

\section{Building certificates and checking them in Lean}
\label{sec:certificates}
\subsection{The derivation graph}
Reaching the desired equation is not enough; the solver must preserve how it got there. Completion records this explanation as a graph of equalities with no cycles. Each node stores an equality, the earlier nodes it uses, and the rule or substitution that justifies it. Several later steps can reuse the same earlier lemma. There are six node kinds: hypothesis instance, reflexivity, symmetry, transitivity, congruence, and substitution. The first five use the equality rules introduced in Section \ref{sec:representation}; the sixth substitutes expressions into an already proved equality. Rewriting inside a larger expression applies congruence along the path through its tree. This is how the searches of Section \ref{sec:true} keep a checkable history.
Before generating Lean code, a local checker tests that every referenced parent exists, the graph has no cycles, and the equations fit together. For transitivity, the first equality's right side must be the second equality's left side. A hypothesis-instance node must contain exactly the input equation after its recorded substitution. These checks catch bookkeeping errors early. The graph records how each equality follows from the hypothesis or from earlier equalities. The Python checker catches mistakes in that record; Lean checks the final proof independently \cite{lean4,lean4lean}. The Python checker itself is not formally verified.

\subsection{Generating Lean code with shared lemmas}
The code generator keeps only the nodes needed for the final equality and places each parent before the nodes that use it. Each node becomes a local \code{have} declaration, Lean code that states and proves an intermediate fact. The generator collects all variables needed by each lemma, including variables used in its supporting derivation but no longer present in its final equation. Substitution nodes apply a parent lemma to new expressions, congruence uses \code{congrArg}, and transitivity chains equality proofs. Reusing lemmas avoids copying the same large derivation each time it is needed.
The generated file imports the judge's problem and defines \code{submission : Goal}, where \code{Goal} is the exact proposition to prove. For a true answer, the certificate takes an arbitrary set and binary operation, assumes the hypothesis, and derives the target. Rewrite-path searches generate short equality chains; completion searches use the shared proof graph. Both prove the same kind of claim, despite the different code shapes.

\paragraph{Completing the running example.}
In Example \ref{ex:running}, the steps that derive $K$ become shared lemmas, and the proof of $P$ uses them. The generated certificate contains only the lemmas needed for the final proof. It applies $P$ to $((a\op b)\op c)\op d$, to $(a\op b)\op c$, and finally to $a\op b$. The judge checks these applications and the equalities that connect them.
The example follows the complete proof process: parse the expressions, derive $K$ and $P$, apply $P$ to the target, and write the recorded steps in Lean. The idempotence-versus-commutativity example uses the counterexample process instead. The solver builds the two-element projection table, checks that idempotence holds and commutativity fails, and expresses those facts in Lean. Appendix \ref{app:evidence} records the two accepted example checks separately from the benchmark results.

\subsection{Certificates for finite and infinite counterexamples}
Finite certificates construct \code{Magma (Fin n)}, an operation on the $n$ values $0,\ldots,n-1$, from a table or modular formula. The official \code{decideFin!} tactic checks every hypothesis assignment and verifies that some target assignment fails. Certificates for larger sets raise local recursion limits. Sets with more than nine elements also receive a larger \emph{heartbeat allowance}, Lean's counter-based limit on computation. These settings allow more work during checking; they do not change the proposition being proved. An infinite certificate cannot check every element by enumeration. Instead, it defines the set by its three forms and proves the base law by a finite case analysis, splitting indices into zero and successor cases where needed. It reconstructs the proof of the input hypothesis from the recorded graph. Finally, the distinct forms show the target violation: $0$ cannot equal $o_n$, and $o_n$ cannot equal $e_m$. In Lean, this last step uses disjointness of the constructors defining the set.

\subsection{What acceptance guarantees}
Before calling the judge, the solver checks that the certificate is nonempty, fits the byte limits, and has the required declaration. It also screens for forbidden proof placeholders and metaprogramming tokens. Some later table searches consult the supplied declaration and axiom policy before they start. This policy lists the Lean definitions and foundational axioms that a submission may use. These checks catch some invalid candidates before submission, but only the official judge can accept an answer. Appendix \ref{app:implementation} states which checks each method uses.
The solver sends the proposed true or false answer with its code. The judge creates the corresponding proposition and checks the code in Lean. Acceptance therefore depends on correct problem translation, imported definitions, allowed axioms, and Lean's proof checking. It does not require trusting the Python search's claim of success. Conversely, a summary saying that search has finished is not a proof. Search can use fallible heuristics because its proposed answers must pass a separate proof check.

\section{Time and size limits}
\label{sec:resources}
The official runner, called the \emph{proxy}, supplies each problem and its resource limits, and mediates certificate checks between the solver and judge. Appendix \ref{app:implementation} records the communication protocol.
The default overall allowance is 3,600 seconds. Each search uses a smaller allowance and explicit limits on its working state. An elapsed-time clock tracks the remaining budget, and some routes reserve time for later methods. The proxy enforces the overall deadline; local checks and process timers regulate individual stages. Appendix \ref{app:parameters} gives the exact bounds and reserves. The general certificate limit is the smaller of the allowance supplied by the runner and 50,000 UTF-8 bytes. Counterexample certificates have an additional limit, 20,000 bytes by default. The solver may therefore find a proof, discard it because it is too large, and continue searching. Limits on proof nodes, queues, and generations help control memory use and proof size before a certificate is built.
The schedule favors quick successes and keeps the early stages that worked in development tests. But it is a fixed list of bounded searches, not an \emph{anytime search} that continues using the available time. Finishing the list does not automatically enlarge a queue or try a larger set. A run with an hour available can therefore stop unsolved after a few minutes. This behavior matters when we examine the missed main-set case.

\section{Evaluation}
\label{sec:evaluation}
\subsection{What was measured and how}
The evaluation records the solver's success rate, the methods that find each answer, the runtime of each case, and the size of accepted certificates. The measurements come from saved run records for the exact submitted file, identified by its hash. Appendix \ref{app:evidence} lists the records and software revisions.

\paragraph{Official result.}
The Solo / OpenAI gpt-oss-120b main leaderboard reports \sys{} at 199/200, tied rank six, with a total elapsed time of 47 minutes 43 seconds. Its sole error is \code{order5_normal_0050}, the same equation pair left unsolved in our local tests. The public page does not provide a full record of certificates or language-model calls from the organizers' run. We therefore do not know whether that run used the optional language-model fallback.

\paragraph{Local conditions.}
Our local reruns used Lean~4.33.1 and Python~3.11.14. We disabled the optional language-model fallback to measure the mathematical searches alone. The main-set rerun used four workers. Appendix \ref{app:optionalproposal} explains how the test runner disabled the fallback; Appendix \ref{app:evidence} records the conditions of earlier qualification runs.

\paragraph{What each suite tests.}
The public qualification set contains 1,000 normal cases and hard collections of 69, 200, and 400 cases. The previous-stage evaluation has four groups of 200 cases. We used these sets during development and qualification. Solving all of them shows that tested successes were preserved, not how well the solver generalizes to unseen problems. The main set is the official 200-case Stage 2 evaluation set, rerun locally after its release. It has fifty cases in each of four categories: normal, hard, and extra-hard order-four equations, and normal order-five equations.

The \emph{long-chain true} test set contains 200 implications selected from a fixed snapshot of the Equational Theories Project graph. In this graph, an edge is an explicitly proved implication between two laws. The selected pairs have shortest paths of fifteen to nineteen edges. Selection excludes public pairs, simple collapse hypotheses, and low-complexity laws. This tests whether the solver can derive an implication without following the graph's known chain.

The \emph{finite/general boundary} test set uses the same snapshot and contains 821 distinct pairs absent from the public sets. These implications fail when infinite magmas are allowed, but hold, or have unknown status, when restricted to finite magmas. For 819 pairs, every finite magma satisfying $H$ also satisfies $T$: only an infinite countermodel can disprove the implication. The finite status of the other two is unknown. This set tests which infinite counterexamples the solver can find, rather than its ability to search finite tables.

\Needspace{7\baselineskip}
Three metrics keep success and reliability separate:
\begin{itemize}
\item \emph{Coverage}: the fraction of cases producing a certificate submitted to the judge.
\item \emph{Certificate acceptance}: accepted attempts divided by submitted attempts.
\item \emph{Solved-case rate}: cases with an accepted certificate divided by all cases.
\end{itemize}
Every certificate submitted to the judge was accepted in the completed local replays. Coverage therefore equals solved-case rate, while certificate acceptance is 100\%. For the main replay these are 199/200 and 199/199, respectively. An unsolved case lowers coverage even though it supplies no incorrect certificate. Timing includes the test runner and judge, not just Python search. We report sums of individual case times. These differ from \emph{campaign elapsed time}, the time from the first case starting to the last case finishing. Parallel workers can reduce elapsed time without reducing the sum. The records do not contain complete hardware details, so these timings describe the saved runs rather than a controlled hardware comparison. Appendix \ref{app:metricnotes} explains the labels and timestamps.

\subsection{How many problems were solved?}
\begin{table}[t]
\centering\small
\begin{tabular}{@{}lrrrrr@{}}
\toprule
Suite & Cases & Accepted & Solved rate & Rejected attempts & Unsolved \\
\midrule
Public qualification & 1,669 & 1,669 & 100.00\% & 0 & 0 \\
Previous-stage evaluation & 800 & 800 & 100.00\% & 0 & 0 \\
Released Stage 2 main & 200 & 199 & 99.50\% & 0 & 1 \\
Long-chain true & 200 & 197 & 98.50\% & 0 & 3 \\
Finite/general boundary & 821 & 147 & 17.90\% & 0 & 674 \\
\bottomrule
\end{tabular}
\caption{Retained qualification and local replay results. All certificates submitted to the judge were accepted; solved rates include unsolved cases in their denominators.}
\label{tab:results}
\end{table}
The public-pair list has 2,669 entries but only 2,469 distinct pairs: 200 legacy entries repeat existing pairs. Both additional test sets exclude every public pair, contain no repeated pairs, and have no pairs in common with each other. Appendix \ref{app:metricnotes} gives the count breakdown. The main replay accepted all fifty cases in each order-four category and 49 of fifty order-five normal cases. Its accepted verdicts were 99 true and 100 false. The remaining pair produced no certificate locally and appears as the sole error in the hosted matrix. Thus the replay matches both the official count and the identity of the missed pair. Appendix \ref{app:mainfailure} analyzes its local trace.

\subsection{Runtime and certificate size}
\begin{table}[t]
\centering\small
\begin{tabular}{@{}lrrrrr@{}}
\toprule
Suite & Median time (s) & 95th pct. (s) & Max. (s) & Median cert. (B) & Max. cert. (B) \\
\midrule
Released main & 2.835 & 9.652 & 186.506 & 255 & 40,292 \\
Long-chain true & 2.857 & 3.487 & 190.370 & 3,700 & 34,260 \\
Boundary & 121.469 & 164.890 & 242.183 & 1,499 & 5,122 \\
\bottomrule
\end{tabular}
\caption{Runtime over all cases; certificate sizes over accepted cases only. Times include search and checking in the recorded local environment. The 95th-percentile values use interpolation, as in the saved summaries.}
\label{tab:runtime}
\end{table}

The sum of case times was 881.958 seconds for the main rerun, 1,241.349 seconds for the long-chain set, and 81,664.970 seconds for the boundary set. Most long-chain cases finished quickly; three unsolved cases took much longer. Boundary cases often ran through the earlier methods before reaching the infinite construction or stopping unsolved, giving a much higher median. The sets therefore test different costs: deriving useful equations, and spending time on unsuccessful methods before trying an infinite model.
The organizers' run and our local rerun are different executions. The displayed official total is 2,863 seconds, about 3.25 times the local sum, but hardware, scheduling, time accounting, and settings were not controlled. This ratio is not a measured speedup. Each local test set has one completed rerun. The median accepted certificate on the main set is only 255 bytes. Many answers therefore have a compact model description or proof, even though the solver includes methods for much larger searches. The largest certificate is 40,292 bytes, showing the range of certificate sizes that the implementation has to handle.

\subsection{Which routes supplied accepted certificates?}
The main replay's false certificates used carrier sizes two (29 cases), three (3), four (45), five (2), seven (8), eight (8), and eleven (5). This distribution is useful in two ways. First, larger structured families matter: accepted witnesses extend well beyond the exhaustive two- and three-element search. Second, most finite witnesses are not five-element models merely because an equation category is called order five. For true answers on the main set, the records attribute 68 cases to other deterministic proving methods, five to one-generation completion, 24 to the multi-generation size/depth configuration, and two to target-interleaved completion. The broad label ``other'' combines the early direct, overlap, projection, and conversion methods. The existing logs cannot separate their contributions more finely. Completion accounts for more successes on the long-chain set: 58 accepted cases used early deterministic proofs, seven used one-generation completion, 131 used multi-generation completion, and one used lazy projection. The graph paths used to select these cases have fifteen to nineteen edges. These distances count known implications, not the minimum number of equational proof steps. The early successes show that the solver need not follow those known graph paths. Every accepted answer on the boundary set used the infinite ray model or its opposite operation. The set contains 110 distinct hypotheses, but all 147 successes came from six of them. The accepted certificates have 70 distinct hashes. These results reflect repeated use of the same model across different targets, not the discovery of many different infinite models. Appendix \ref{app:boundary} breaks down these successes by hypothesis and distinguishes an unsuitable model from failures to prove the hypothesis or find a target violation. The long-chain and boundary sets were selected differently, so their 197/200 and 147/821 results do not directly compare the general difficulty of the two kinds of problem.
These counts identify the first successful method in the fixed schedule. They do not show how many cases depend on that method: another might solve the same problem if run alone or given more time. The evaluation did not include \emph{ablation} tests, which disable or reorder components, so it does not isolate each component's added benefit.

\section{Reproducing the results}
\label{sec:repro}
The public repository\footnote{\url{https://github.com/hieronimoDos/rozsa-equational-solver/tree/paper-v1}} contains the unchanged solver, individual run records, additional test sets, graph snapshots, and the scripts needed to inspect or rerun the experiments. The release for this manuscript is \code{paper-v1}, at Git commit \code{df735b4}. Appendix \ref{app:implementation} gives the solver's SHA-256 hash; this identifies the submitted file even though its internal version label is outdated. The solver and author-created code are Apache-2.0 licensed, with third-party attribution recorded separately.

\paragraph{Artifact availability.}
The repository's \code{README.md} gives the setup commands. To obtain the organizer datasets, use \code{tools/fetch_data.py}. It downloads a fixed revision and verifies each checksum; the repository does not redistribute those datasets. The script \code{tools/extract_evidence.py} recomputes the reported statistics from saved records. It does not run the solver or contact a judge. The script \code{tools/build_suites.py} rebuilds the additional test sets from the supplied graph snapshots. The published records replace machine-specific path prefixes with placeholders, but leave numerical results, proof code, and certificate hashes unchanged. The provenance inventory lists both the original and published file hashes. To run new checks, obtain the official challenge repository\footnote{\url{https://github.com/SAIRcompetition/equational-theories-lean-stage2}} at the revision recorded for the relevant experiment and build its Lean toolchain and dependencies. After fetching the datasets, run the following from the release checkout:
\par\noindent\begin{minipage}{\linewidth}
\begin{lstlisting}
python3.11 tools/replay.py --official official-judge \
  --submission submission \
  --problems data/official/stage2_evaluation_main.jsonl \
  --output new_results/main.json --timeout 3600
\end{lstlisting}
\end{minipage}
Here \code{official-judge} is the built judge checkout, and \code{submission} is the directory containing the supplied solver. The replay script disables language-model calls, runs the solver, and checks its certificates through the official runner. It handles one case at a time, rather than using the four workers of the original main-set rerun. Its total elapsed time is therefore not a reproduction of that run's elapsed time. The output path must name a new result file. Record the Python version, process allowances, worker count, judge revision, dataset hash, and certificate hashes. Reproducing a result means both finding an answer and obtaining Lean's acceptance, so the runner and judge setup are part of the experiment.

\section{Discussion and conclusion}
\label{sec:discussion}
\subsection{What the result establishes}
The submitted solver achieved 199/200 in the official evaluation. A local rerun using only the mathematical methods solved the same cases, and development checks preserved all earlier tested successes. The additional sets show where this approach works well and where it is limited. The prover solves most of the selected implications connected by long known chains, without following those chains. In contrast, its infinite counterexamples depend on a narrow fixed construction. What makes the methods fit together is the requirement to produce a checked certificate. A proof search records why each new equation follows. A model search records an explicit operation and a target violation. Both must turn their result into a proof the same judge accepts. This lets a small program combine different kinds of mathematical search without trusting the search heuristics themselves.

\subsection{Lessons from post-evaluation analysis}
The missed main-set case failed locally because search found no answer, not because Lean rejected a proof. The log shows that internal limits ended the searches while much of the overall time allowance remained. Some queues still contained unprocessed work, but that does not establish that they contained a proof. One attempted projection lemma cannot follow from the hypothesis at all. Appendix \ref{app:mainfailure} proves this with a two-element model and separates the impossible intermediate goal from searches that stopped at their configured limits. This distinction explains the failure more precisely than a shortage of total runtime. The solver spent part of its search on an intermediate goal that was mathematically impossible, while other bounded searches ended without deciding the original implication. The post-evaluation calculation rules out that intermediate goal; it does not determine the original implication's truth value or show that the remaining searches would have succeeded.

\subsection{Limits of the evidence}
The boundary cases also separate limits of the chosen model from limits of the search. Appendix \ref{app:boundary} describes four possibilities: the model fails the hypothesis, it satisfies both equations, the search misses an assignment where the target fails, or the search cannot prove the hypothesis. Testing finitely many indices can find a violation, but cannot establish an identity on the whole infinite set without a proved cutoff or an argument covering every index. The records count successes for each hypothesis; they do not classify every failure by these reasons. The measurements cover fixed benchmarks and one completed local rerun per set. They do not establish run-to-run variability, isolate each method's added coverage, or estimate performance on an arbitrary new distribution. Hardware and execution conditions were not fully controlled, so the official and local runtimes are not a measured speed comparison. The records contain certificate byte lengths, not counts of equational proof steps.

\sys{} brings proof search and counterexample search into one program, with Lean checking every submitted answer. The searches use different methods, but each records enough evidence to build a certificate. The submitted solver achieved 199/200 on the official main evaluation, and the local mathematical searches solved the same 199 cases. They also solved 197/200 selected long-chain implications and 147/821 boundary cases. The latter result makes the limited reach of the infinite-model search clear. The completed system thus combines high coverage on the main benchmark with independently checked answers and a documented account of the cases it did not solve.

\paragraph{Acknowledgments and preparation.}
We thank the Equational Theories Project and challenge organizers for the mathematical corpus and certificate-checking infrastructure. AI assistants supported source inspection, code writing, and manuscript preparation. Responsibility for the manuscript rests with the author.

\appendix
\section{Search order and main limits}
\label{app:parameters}
\subsection{Exact runtime order}
\label{app:order}
The function \code{run()} tries the methods in the following order. Every proposed answer passes the same local checks and judge call. An accepted answer ends the run.
\begin{enumerate}
\item Structured models at sizes 2--5; exhaustive tables at sizes 2 and 3.
\item Direct and reversed instances; total collapse; recorded breadth-first search (BFS), depth 4, 3,500 states.
\item Seeded random tables at sizes 4 and 5, 256 attempts per size.
\item Extended structured models at sizes 6--9, including switched-graph tables at size 9; binary affine models if the table policy permits them.
\item Up to 16 self-overlap conversion candidates; self-overlap projection candidates; BFS-derived projection lemmas.
\item Conditional early lazy projection completion, 45 seconds per applicable direction.
\item The guarded seven-entry recording-completion portfolio.
\item Joint structural conversion; active completion.
\item Modular affine search at sizes 10--17.
\item Deeper recorded BFS, depth 8, 11,000 states; shrinking reverse conversion; structural and guided conversion.
\item Remaining binary affine route if not attempted; rank-one quadratic and gated unary routes.
\item Violation-directed local repair at sizes 4, 5, 6; joint local repair at the same sizes.
\item Infinite ray certificate search; six-element constraint satisfaction (CSP); late lazy projection completion.
\item The optional proposal interface described in Appendix \ref{app:optionalproposal}; otherwise log unsolved.
\end{enumerate}

\subsection{Completion portfolio}
\begin{table}[ht]
\centering\small
\begin{tabular}{@{}lrrrrrr@{}}
\toprule
Entry & Seconds & Generation & Rules & Pairs & Proof nodes & Queue \\
\midrule
P0 & 0.05 & 0 & 16 & 100 & 3,000 & 200 \\
P1 & 0.12 & 1 & 32 & 350 & 7,000 & 800 \\
P2 & 1.25 & 6 & 96 & 2,200 & 25,000 & 4,500 \\
P3 & 1.00 & 5 & 112 & 1,800 & 24,000 & 4,000 \\
P4 & 1.00 & 6 & 96 & 1,800 & 24,000 & 4,000 \\
P5 & 2.00 & 10 & 144 & 3,500 & 40,000 & 7,000 \\
P6 & 12.00 & 15 & 96 & 10,000 & 100,000 & 15,000 \\
\bottomrule
\end{tabular}
\caption{Internal search limits. A counter can be just above its cap when the check occurs. The seconds column is an elapsed-time check, separate from a 30-second process timer for each configuration. Other limits control overlaps, equations, term size/depth, and normalization steps.}
\end{table}
P0--P2 and P4 use the same ranking by size and then depth. P3 uses unit-weight term size and a lexical comparison of printed expressions. P5--P6 rank by depth and then size. P4 keeps the source label \code{lpo}, but does not implement a separate lexicographic path order. P6 alternates overlap generation with target-focused processing. Its maximum term size/depth is 61/18 and its normalization limit is 80 steps. The other configurations range from size/depth 31/10 to 51/15 and 20 to 56 normalization steps. The final function alias in the file selects the guarded completion function used at runtime.

\subsection{Further major limits}
Active completion has a two-second internal budget, an eight-second process timer, a 60,000-node bound, 192 generators, 12,000 critical pairs, and term size/depth 81/23. Lazy projection completion defaults to 25 seconds per applicable projection, with a five-second extra timer margin, approximately 100,000 proof nodes, 320 generators, 180,000 pairs, and 90,000 queued entries. These are checked thresholds, not exact allocation guarantees. The extended structured pass receives at most twelve seconds. Modular affine search receives at most five seconds. Quadratic and gated unary searches share a ten-second deadline. Local-repair allowances at sizes 4/5/6 are 20/30/75 seconds; joint-local allowances are 20/35/100 seconds with inspection limits 24,000/30,000/50,000. The six-element CSP receives fourteen seconds and at most 50,000 nodes. The search that derives the hypothesis from the ray model's base law receives 0.2 seconds for each of the original and opposite operations, four generations, 400 critical pairs, and 4,000 proof nodes. A stage may be skipped if too little time remains or if that time is reserved for later methods.
Many proof searches require more than 120 seconds to remain. Other stages reserve 35 or 90 seconds for later work. These settings limit each search and preserve time for later methods; they are not claimed to allocate the budget optimally.

\subsection{Rewrite-search limits, entry conditions, and queue priorities}
\label{app:conversionbounds}
The initial breadth-first conversion admits path depth four and 3,500 states; the later pass admits depth eight and 11,000 states. Guided conversion allows 12,000 states and path depth forty. Its distance is zero for identical subtrees, one for unequal variable leaves, the combined sizes for different node kinds, and the sum of recursive distances for operation nodes. P6 scores a queued equality by the total size of its two sides, minus a count of alphabetic characters shared with the printed target. Depth, generation, and age break ties. Renaming variables can therefore change this score. Local table repair normalizes problem descriptions when choosing seeds, but that normalization does not apply to proof-search scores. Active completion orients $s=t$ as $s\to t$ only if each variable occurs at least as often in $s$ as in $t$ and $(\operatorname{size}(s),\operatorname{depth}(s))$ is lexicographically greater than the corresponding pair for $t$. It tests the reverse direction similarly. Both guards can fail, leaving the equation available as a bidirectional overlap generator. The early projection search runs only if at least two target variables are absent from the hypothesis. The hypothesis's left side must be a variable equal to the target's left side, and that variable must occur at least twice on the hypothesis's right side. The target sides must also have matching leftmost leaves or matching rightmost leaves. Each applicable projection direction then receives 45 seconds. The late projection search uses the 25-second default.

\section{Infinite ray operation and finite certificate example}
\label{app:ray}
For the set with elements $0,o_n,e_n$ introduced in Section \ref{sec:false}, the following table defines the ray operation everywhere. The $o_0$ and $o_{n+1}$ cases must be separate.
\begin{center}
\begin{tabular}{@{}lccc@{}}
\toprule
Left operand & Right $0$ & Right $o_m$ & Right $e_m$ \\
\midrule
$0$ & $o_0$ & $0$ & $o_0$ \\
$o_0$ & $0$ & $e_0$ & $0$ \\
$o_{n+1}$ & $e_n$ & $e_{n+1}$ & $e_n$ \\
$e_n$ & $o_{n+1}$ & $o_n$ & $o_{n+1}$ \\
\bottomrule
\end{tabular}
\end{center}
The right input's index $m$ never affects the result; only its form matters. The left input's index $n$ does affect the output, and $o_0$ must be treated separately from $o_{n+1}$. To prove~\eqref{eq:raylaw}, the code considers the three possible forms of each of $x,y,z$, then splits indices into zero and successor cases where needed. To show that the target fails, it checks that the two results have different forms, such as $0$ and $o_n$. The submitted search does not certify inequalities between elements of the same form with different indices. For a finite table, the generated certificate has the following form. The table here is the left projection operation from the first worked example. The judge-generated \code{Goal} states the equations and quantifiers to be proved.
\par\noindent\begin{minipage}{\linewidth}
\begin{lstlisting}
import JudgeProblem
import JudgeDecide.DecideBang
import JudgeFinOp.MemoFinOp
open MemoFinOp
def submission : Goal := by
  let m : Magma (Fin 2) := { op := finOpTable "[[0,0],[1,1]]" }
  refine (*@$\langle$@*)Fin 2, m, ?_(*@$\rangle$@*)
  decideFin!
\end{lstlisting}
\end{minipage}
The code supplies the two-element set and its operation table, then asks the finite checker to verify that the hypothesis holds and the target fails. The angle brackets are Lean syntax for supplying these parts of the counterexample together.

\section{Source map and saved run records}
\label{app:evidence}
\subsection{Source map}
Line numbers refer to the unchanged 4,074-line solver, not to a reformatted version.
\small
\begin{longtable}{@{}p{.25\linewidth}p{.55\linewidth}r@{}}
\toprule Component & Representative live functions & Lines \\
\midrule\endhead
Term operations & \code{parse_term}, \code{match}, \code{substitute} & 17--230 \\
Protocol/classification & \code{call_judge}, \code{call_llm}, \code{classify_equations} & 232--311 \\
Direct/collapse & \code{direct_proof_bodies}, \code{collapse_proof_body} & 312--371 \\
Small finite search & \code{structured_tables}, \code{search_exhaustive} & 374--509 \\
Recorded conversion & \code{recording_bfs}, \code{guided_recording_body} & 518--760 \\
Projection/overlap & \code{derived_projection_proofs}, \code{critical_conversion_proofs} & 763--1215 \\
Finite emission/policy & \code{make_false_code_any};\newline\code{proof_policy_allows_false_table} & 1218--1304 \\
Further models & affine, quadratic, gated and local search functions & 1307--1624 \\
Certificate/LLM screen & \code{certificate_compliant}, \code{parse_llm_candidate} & 1684--1860 \\
Orchestration & \code{run} & 1905--2205 \\
R2 terms/unification & \code{r2_parse_term}, \code{r2_unify} & 2230--2561 \\
Derivation graph & \code{r2_DerivationArena} & 2603--2762 \\
Rankings/critical pairs & \code{r2_orient}, \code{r2_generate_critical_pairs} & 2766--2915 \\
Proof emission & \code{r2_emit_tactic}, \code{r2_make_true_code} & 2968--3102 \\
Completion/guard & \code{r2_complete}, \code{r2_try_r2_completion_guarded} & 3173--3359 \\
Active completion & \code{r4_active_complete} & 3372--3532 \\
Lazy projection & \code{r5_lazy_projection_complete} & 3535--3733 \\
Six-element CSP & \code{r6_FiniteMagmaCSP}, \code{r6_csp_countermodel} & 3736--3948 \\
Infinite ray & \code{r7_ray_certificate} & 3951--4068 \\
Live hook and entry & guarded R2 alias and \code{main} & 4071--4074 \\
\bottomrule
\end{longtable}
\normalsize

\subsection{Record identifiers and hashes}
The file \code{evidence/official_leaderboard_20261008.json} records the official main-set result: Solo / OpenAI gpt-oss-120b, team EQT02-T00223, tied rank six, 199/200, 171.265~KiB, and 47 minutes 43 seconds. Its sole error is $(\E{26286},\E{20170})$, \code{order5_normal_0050}. This saves the public leaderboard and per-problem result matrix, not a complete log of the organizers' execution.
The accompanying \code{evidence/paper_evidence.json} lists the source records and compact main-case summaries. Its named entries are:
\begin{itemize}
\item \code{public1669}: public qualification counts, published-label agreement, per-case certificates, and strategy events.
\item \code{stage1_800}: previous-stage acceptance, label agreement, certificate hashes, and runtime.
\item \code{qualification} and \code{lean4331_targeted}: release identity, preserved certificates, and targeted toolchain qualification.
\item \code{main}: the completed released-main summary; \code{miss}: the full residual-case trace.
\item \code{boundary_report}: coverage, certificate acceptance, per-strategy metrics, and time aggregation for both additional suites.
\item \code{boundary_manifest}: dataset construction, exclusions, source hashes, and distinct-pair counts.
\end{itemize}
The qualification and local rerun records report zero actual LLM calls. This describes those saved runs, not the organizers' run, for which no call log is available. The two teaching examples were checked separately and are recorded in \code{evidence/illustrative_examples.json}. The unchanged completion method produced an accepted certificate for the running overlap example. The judge also accepted the finite projection example under the official runner's default proof policy. Neither check is included in the benchmark totals.

The main, boundary, and long-chain dataset hashes are respectively:
\begin{quote}\footnotesize\ttfamily
10eb23ddba875305c7b54f08f5e742261fb3bde5a32e413add3199f702960756\\
6426a1732fffa5c19b5a70c5a9560d64f4cc175cb5fbfbafa1a896a79de1cdf0\\
96017f5711570324224e0310a916e6edfa486102ee95a73b5177840d3ce64443.
\end{quote}
The boundary graph records separately whether an implication holds for all magmas and whether it holds for finite magmas. The long-chain set uses shortest paths in the graph of explicitly proved implications, choosing long paths and excluding public pairs and direct edges. Both laws have at least four operation occurrences. A hypothesis is excluded if the unrestricted graph says it implies $x=y$, the law forcing a nonempty magma to have one element.
Targets have the label \code{implicit_proof_true}. Candidate distances are at least three; the selected cases have distances fifteen to nineteen. Selection prefers larger distances, allows each hypothesis and each target to appear at most twice, and breaks ties using a fixed hash. The set-building script and manifest record these rules.

\subsection{Metric and timestamp conventions}
\label{app:metricnotes}
The public qualification records include published true/false labels, so agreement with those labels can be counted directly. The released-main records contain no reference verdict; for those cases, success is measured by the official certificate judge's acceptance. An empty ``correct accepted'' field in a generic summary is not treated as a correctness count when reference labels are absent. The public inventory consists of 1,669 qualification entries, 800 previous-stage entries, and 200 legacy \code{hard.jsonl} entries repeating pairs already in \code{hard1.jsonl}. Thus 2,669 entries represent 2,469 distinct pairs. All are included in the exclusion audit for the boundary and graph-chain suites. The runner for the additional test sets can use parallel workers and resume partial results. Campaign elapsed time is measured from the first case's start to the last case's finish, not from timestamps on the combined report. Summed case time adds the duration of every case. This distinguishes time accumulated across parallel processes from time passing on the clock.

\subsection{Implementation and environment notes}
\label{app:implementation}
The described submission is a regular read-only file of 175,375 bytes, identified by SHA-256:
\begin{center}\small\ttfamily
c98f50b78462e6305cf0518f475df2bea0efce82e02d5a0153ef82f5f79a5402.
\end{center}

At startup, the official proxy supplies the problem and budget through a line-oriented JSON protocol. Standard output carries protocol messages; standard error carries diagnostics. Some source labels describe earlier versions rather than current behavior. The file's \code{VERSION} string still says R1, although later completion and model methods are included. Its \code{COMPLETION_STATUS} string is obsolete: \code{run()} does invoke completion. The final alias selects \code{r2_try_r2_completion_guarded}. A planned e-graph, a structure for sharing known equalities, was not implemented. \code{JointConstraintState} and its \code{nogoods} field are unused. The running solver has no SAT, SMT, or conflict-driven clause-learning backend; these names refer to general constraint-solving techniques, not hidden components of this submission. Several rule-ordering names overstate what their implementations do. The \code{lpo} comparison gives the same results as \code{size_depth}: its special subexpression cases agree with the size/depth fallback. The \code{rpo} comparison similarly agrees with depth/size and is not used by the current search sequence. The \code{kbo} comparison uses unit-weight term size and printed lexical order, but does not enforce every condition of a standard Knuth--Bendix order. Because each comparison ranks every pair of distinct terms, the R2 \code{unfailing} branch for equations that cannot be oriented never runs. The seven configurations still differ in their limits and in when they process equations resembling the target. The helper used to choose active rewrite directions has a name suggesting that substitutions preserve its ranking. They need not: when sizes tie, replacing variables by expressions can reverse the depth comparison. For instance, $(((a\op b)\op c)\op d)$ and $(a\op c)\op(b\op d)$ have size/depth pairs $(7,3)$ and $(7,2)$. Substituting $d=(((u\op u)\op u)\op u)$ gives $(13,4)$ and $(13,5)$. The variable-occurrence test holds in the original terms. This is why the paper does not use that ranking to claim termination; bounded rewriting with recorded proofs is sufficient for the accepted paths. Joint local search chooses a seed from a normalized problem description. Simultaneously renaming variables in the pair, swapping equation sides, or mirroring both input equations leaves that description unchanged. Renaming the separately quantified hypothesis and target variables independently can change it. These are the transformations that preserve the random-number stream, but set iteration and time cutoffs can still affect the completed search. The proxy's default policy permits \code{Classical.choice}, \code{Quot.sound}, and \code{propext}, with specified declarations and prefixes. Selected late-table routes check policy compatibility; earlier table routes and the LLM table path rely on the official certificate check. Calling the low-level judge with an omitted or stricter policy changes the experiment. The pedagogical checks use the proxy's policy. Full judge, Lean, mathlib, and solver revision hashes are retained in the evidence inventory and README.

\subsection{Optional language-model fallback, disabled in local tests}
\label{app:optionalproposal}
The submitted file also contains an optional large language model (LLM) fallback, tried after the mathematical searches. It asks for a proposed answer to a remaining problem, not for a result the solver can trust without checking. Every proposal still needs an accepted certificate. The local benchmarks disable this fallback, and no experiment here measures whether it solves additional cases. Its use in the official evaluation is unknown because the public record contains no model-call log. The local results therefore measure the mathematical searches, not the benefit of this extension. If enough time remains, the fallback allows at most four rounds of proposals and feedback. The first asks for analysis; later rounds ask for repair of a true or false candidate. The supplied context includes hypothesis argument order, normalized equations, labels based on syntax, search summaries, nearby rewritten expressions, the previous attempt and judge message, and remaining time. It does not include unfinished search queues. A true proposal gives a Lean proof body; a false proposal gives an integer operation table. Tables receive full local validation, and proof bodies receive the required submission wrapper. Both then pass the local certificate checks and the judge. Feedback can help repair a proposal, but cannot replace proof checking. The official runner reads the top-level \code{PROMPT} string from the Python source without executing the file. It fills the placeholders with context and judge history, then uses the model and provider configured by the organizers. The prompt asks for a complete Lean proof body or square integer table and discourages unsupported theorem names and proof placeholders. Our local test runner intercepts these requests and returns a disabled-fallback response without contacting a model. This tests the mathematical searches without changing the submitted file.

\section{Post-evaluation failure analysis}
\label{app:failures}
This appendix separates facts in the saved logs from later mathematical deductions about the completed solver. None of the analysis changes the submission. ``Unsolved'' means no accepted certificate was found, not that the implication is false. ``Rejected'' means the judge received and rejected a certificate. Every completed local test set reported here has zero rejected attempts, including the unsolved cases below.

\subsection{The missing main-set point and its search trace}
\label{app:mainfailure}
The missing local row is \code{stage2_evaluation_main_0200}. The hosted problem identifier is \code{order5_normal_0050}. Both refer to $(\E{26286},\E{20170})$:
\begin{align}
H:\quad x &= (y\op((z\op x)\op z))\op(y\op x),\label{eq:missH}\\
T:\quad x &= (y\op z)\op((x\op(w\op z))\op x).\label{eq:missT}
\end{align}
The local rerun, with the LLM disabled, finished after 186.506 seconds. It made zero judge calls, produced zero certificates, and reached one intercepted LLM request. It recorded no infrastructure error and reported 3,415.84 seconds left in the overall allowance. The organizers' result matrix marks the same pair as an error, but does not expose the full execution log.

\begin{table}[ht]
\centering\small
\begin{tabularx}{\linewidth}{@{}l X@{}}
\toprule
Recorded route & Local residual-case outcome \\
\midrule
Exhaustive finite search & Completed sizes two and three; no separating table. \\
Completion P2--P5 & Reached configured critical-pair limits. \\
Target-interleaved P6 & Reached approximately 100,000 proof nodes; 6,644 pairs remained queued. \\
Active completion & No target derivation within its allowance. \\
Local table searches & No witness at sizes four, five, or six; bounded heuristic searches. \\
Joint local searches & Reached inspection caps of 24,000, 30,000, and about 50,000. \\
Six-element CSP & No witness in 14.002 seconds; 3,297 visited search nodes. \\
Late right-projection search & No proof in 25.280 seconds; 50,416 pairs remained queued. Its subgoal is impossible, as proved below. \\
\bottomrule
\end{tabularx}
\caption{Retained local stopping diagnostics. A queue left unfinished does not establish that it contains a useful proof. The projection obstruction is a separate post-evaluation deduction.}
\label{tab:miss}
\end{table}

The early 45-second projection search was skipped because its entry condition requires at least two target variables that do not occur in the hypothesis. Here the hypothesis uses $x,y,z$ and the target uses $x,y,z,w$, so only $w$ is new. The later projection search uses the 25-second default. Only right projection passes the target's variable test: both sides end in $x$, while their leftmost variables differ because the right-hand side begins with $y$.

\subsection{An impossible intermediate law}
\label{app:projectionobstruction}
Let $\sigma:G\to G$ be a nonidentity involution, so $\sigma(\sigma(a))=a$, and define $a\op b=\sigma(b)$. The right-hand side of~\eqref{eq:missH} is
\[
(y\op((z\op x)\op z))\op(y\op x)
=\sigma(y\op x)=\sigma(\sigma(x))=x.
\]
The right-hand side of~\eqref{eq:missT} is
\[
(y\op z)\op((x\op(w\op z))\op x)
=\sigma((x\op(w\op z))\op x)
=\sigma(\sigma(x))=x.
\]
Thus this entire family satisfies both $H$ and $T$. It does not satisfy right projection $R:a\op b=b$, since that law would require $\sigma$ to be the identity. Taking $G=\{0,1\}$ and $\sigma(b)=1-b$ gives the concrete table
\[
\begin{pmatrix}1&0\\1&0\end{pmatrix}.
\]
Consequently $H\not\sat R$. No sound proof search, regardless of its queue cap or time allowance, can prove this particular intermediate law from $H$. This does not settle $H\sat T$: refuting a sufficient intermediate law is not refuting the original target. As a separate post-evaluation check, enumerating all sixteen two-element and 19,683 three-element tables finds two and four labelled tables satisfying $H$, respectively. Every such table also satisfies $T$; only one table at each size satisfies right projection. This confirms that neither size contains a countermodel to the original implication, while a small model does refute the intermediate projection law. The symbolic involution calculation applies at arbitrary carrier size; the finite enumeration does not prove the general implication. The script \code{tools/check_posthoc.py} reproduces the symbolic and finite checks without running or modifying the solver.

\subsection{What the obstruction and unfinished searches establish}
The submitted finite search looked for countermodels to the original implication; it did not retain models to test intermediate lemmas. Model-guided proving uses models to constrain proof search \cite{scott}. A model satisfying $H$ but violating an intermediate law $L$ establishes $H\not\sat L$, whereas agreement with finitely many models does not prove $H\sat L$. The preceding involution calculation rules out the projection law in this way, without deciding the original implication. The remaining logs describe searches that stopped without an answer. P6 left equations in its target-focused queue. Table search exhausted sizes two and three, tried bounded heuristics at sizes four and five, and reached a limit in the six-element constraint search. None of this establishes whether a five-element countermodel exists, whether an infinite countermodel is necessary, or whether $H$ implies $T$. The equation's order does not determine the size of a countermodel.

\subsection{Boundary coverage by source hypothesis}
\label{app:boundary}
The 821 boundary pairs contain 110 distinct hypothesis laws. All 147 accepted answers come from six of those hypotheses: $6/110=5.45\%$ had at least one solved target. This counts hypotheses for which the solver found an answer, not hypotheses that hold in the ray model. The other 104 hypotheses account for 670 pairs, all unsolved. Four more pairs remain unsolved under the six hypotheses that did produce answers.

\begin{table}[ht]
\centering\small
\begin{tabular}{@{}lrrr@{}}
\toprule
Source hypothesis & Pairs & Accepted & Unsolved \\
\midrule
\E{1167} & 36 & 35 & 1 \\
\E{1630} & 2 & 2 & 0 \\
\E{1659} & 37 & 36 & 1 \\
\E{1884} & 2 & 2 & 0 \\
\E{2000} & 37 & 36 & 1 \\
\E{2473} & 37 & 36 & 1 \\
Other 104 sources & 670 & 0 & 670 \\
\midrule
Total & 821 & 147 & 674 \\
\bottomrule
\end{tabular}
\caption{Pair coverage grouped by distinct source hypothesis. Acceptance came exclusively from the fixed ray construction or its opposite; 70 distinct certificate hashes served the 147 accepted pairs.}
\label{tab:boundaryhypotheses}
\end{table}

\Needspace{12\baselineskip}
For the ray model to give a counterexample, $H$ must hold for every assignment and $T$ must fail for at least one. The solver must also find the target violation and prove both claims in Lean. For the original operation and its reverse, there are four possible reasons for failure:
\begin{enumerate}
\item \emph{The operation does not satisfy the hypothesis.} A concrete assignment violates $H$ in the ray magma. Giving the hypothesis-proof search more time cannot change that fact.
\item \emph{The operation satisfies both laws.} If the ray magma satisfies $H$ and $T$ everywhere, it is not a countermodel to this implication.
\item \emph{Search misses a target violation.} The operation satisfies $H$ and violates $T$, but the five-element assignment pool or the test for different constructors misses a witness. Unequal indices within the same constructor are an explicit omitted case.
\item \emph{Search cannot prove the hypothesis.} A separating operation and usable target assignment exist, but the short completion search from the base law fails to derive $H$.
\end{enumerate}
A search may fail both to find a target violation and to prove the hypothesis, so these reasons can overlap. The original and reversed operations must be assessed separately. The records do not classify all 674 unsolved pairs by these reasons.
Testing finitely many indices can find a counterexample to an identity. It cannot prove the identity on the whole infinite set without a proved cutoff or an argument covering every index and element form. The run logs alone therefore do not determine all the identities that hold in this model.
Each of the six successful hypotheses occurs in several target pairs. Solving many such pairs does not mean that the solver found many different infinite models. The 819 implications known to hold in every finite magma have no finite countermodel, regardless of how long finite-table search runs. For those cases, the submitted solver depends on the ray model or its reverse, its limited target assignments, and its search for a proof of the hypothesis.

\subsection{What the different failures mean}
Three outcomes must be kept separate. A \emph{search miss} produces no certificate within the available methods and limits. A \emph{local-screen rejection} finds a candidate that fails a format, policy, or size check before submission to the judge. A \emph{judge rejection} submits code that the official checker does not accept. The missed main case is the first kind; it reached neither of the other two. Local certificate acceptance is 100\%, but coverage is 199/200. The boundary set has the same distinction, at much lower coverage.
The long-chain set's 58 early successes show a different distinction: a long chain of known implications need not require the solver to follow that chain. The records identify the successful routes and certificate byte lengths, but do not count proof steps or establish run-to-run variability. These are limits of the reported evidence, as discussed in Section \ref{sec:discussion}.

\end{document}